\documentclass[twocolumn,superscriptaddress,prb]{revtex4}

\usepackage{graphicx}

\begin{document}

\title{Supercollision cooling in undoped graphene}

\author{A. C. Betz}
\altaffiliation{These authors contributed equally to this work}
\affiliation{Laboratoire Pierre Aigrain, ENS-CNRS UMR 8551,
Universit\'es P. et M. Curie and Paris-Diderot,
24, rue Lhomond, 75231 Paris Cedex 05, France}

\author{S. H. Jhang}
\altaffiliation{These authors contributed equally to this work}
\affiliation{Laboratoire Pierre Aigrain, ENS-CNRS UMR 8551,
Universit\'es P. et M. Curie and Paris-Diderot,
24, rue Lhomond, 75231 Paris Cedex 05, France}

\author{E. Pallecchi}
\affiliation{Laboratoire Pierre Aigrain, ENS-CNRS UMR 8551,
Universit\'es P. et M. Curie and Paris-Diderot,
24, rue Lhomond, 75231 Paris Cedex 05, France}
\affiliation{Laboratoire de Photonique et Nanostructures,
CNRS-UPR20 CNRS, Route de Nozay, 91460 Marcoussis
Cedex, France.}

\author{R. Feirrera}
\affiliation{Laboratoire Pierre Aigrain, ENS-CNRS UMR 8551,
Universit\'es P. et M. Curie and Paris-Diderot,
24, rue Lhomond, 75231 Paris Cedex 05, France}

\author{G. F\`eve}
\affiliation{Laboratoire Pierre Aigrain, ENS-CNRS UMR 8551,
Universit\'es P. et M. Curie and Paris-Diderot,
24, rue Lhomond, 75231 Paris Cedex 05, France}

\author{J.-M. Berroir}
\affiliation{Laboratoire Pierre Aigrain, ENS-CNRS UMR 8551,
Universit\'es P. et M. Curie and Paris-Diderot,
24, rue Lhomond, 75231 Paris Cedex 05, France}

\author{B. Pla\c{c}ais}
\email{bernard.placais@lpa.ens.fr}
\affiliation{Laboratoire Pierre Aigrain, ENS-CNRS UMR 8551,
Universit\'es P. et M. Curie and Paris-Diderot,
24, rue Lhomond, 75231 Paris Cedex 05, France}

\maketitle

\textbf{Carrier mobility in solids is generally limited by electron-impurity or electron-phonon scattering depending on the most frequently occurring event.
Three body collisions between carriers and both phonons and impurities are rare; they are denoted supercollisions (SCs) \cite{Song2012PRL}. Elusive in electronic transport they should emerge in relaxation processes as they allow for large energy transfers \cite{Clarke1991CPL}.
As pointed out in Ref.~\onlinecite{Song2012PRL}, this is the case in undoped graphene where the small Fermi surface drastically restricts the allowed phonon energy in ordinary collisions.
Using electrical heating and sensitive noise thermometry we report on SC-cooling in diffusive monolayer graphene.
At low carrier density and high phonon temperature the Joule power $P$ obeys a $P\propto T_e^3$ law as a function of electronic temperature $T_e$.
It overrules the linear law expected for ordinary collisions which has recently been observed in resistivity measurements \cite{Efetov2010PRL}.
The cubic law is characteristic of SCs and departs from the $T_e^4$ dependence recently reported for  metallic graphene \cite{Betz2012PRL} below the Bloch-Gr\"{u}neisen temperature.
These supercollisions are important for applications of graphene in bolometry and photo-detection.}

\begin{figure*}[t]
\includegraphics[width=17.5cm]{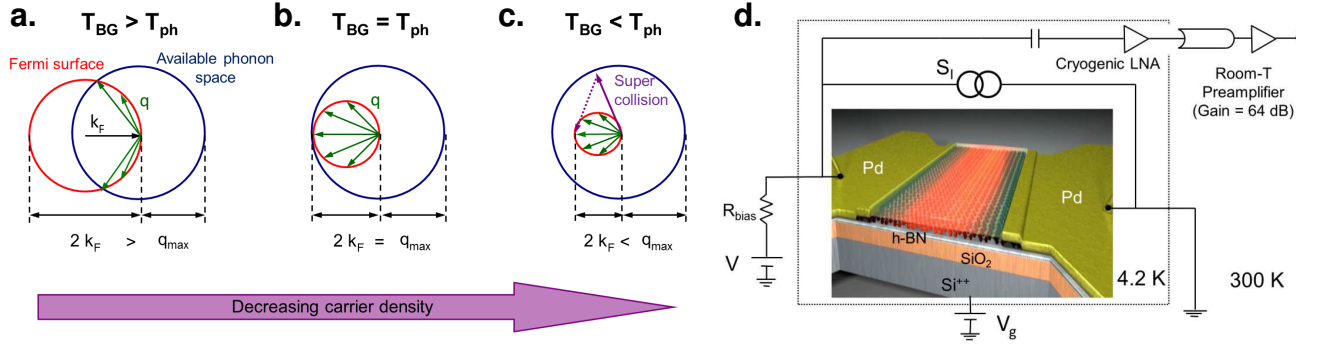}
\caption{ \textbf{Tunability of the Bloch-Gr\"{u}neisen temperature and noise thermometry setup.} \textbf{a},
Electron-phonon interactions scatter carriers from one point on the Fermi surface (red circle) to another, within the boundary of the available phonon space (blue circle).
In low-temperature regime ($T_{ph} < T_{\text{BG}}$), $q_{\text{max}}$ is smaller than 2$k_F$ which represents a full backscattering of electrons.
 \textbf{b},
The Fermi surface shrinks as the carrier density decreases, resulting in a smaller value of $T_{\text{BG}}$. Here, when $T_{ph} = T_{\text{BG}}$, $q_{\text{max}}$ just equals $2k_F$.
 \textbf{c}, In the vicinity of the charge neutrality, one enters the high-temperature regime, where $T_{ph} > T_{\text{BG}}$.
 Here, only phonons with $q \leq 2 k_F$ can scatter off the electrons in the ordinary collisions (green arrows),
 whereas the entire thermal distribution of phonons is allowed for disorder-assisted supercollisions (purple arrow).
 \textbf{d}, Sketch of our noise thermometry setup. Measurements are carried out in liquid helium immersion: A Joule power is supplied to the sample, creating a hot carrier population.
 The hot electrons (diffuse red) induce a thermal noise (current generator $S_I$) and are cooled by the graphene lattice (small balls).
 The current fluctuations are amplified using a cryogenic low noise amplifier (LNA).
 }\label{fig:illustration}
\end{figure*}

Understanding how two-dimensional (2D) electrons in graphene relax their energy to the lattice is not only a central problem in condensed matter physics but also an important issue in the design of graphene devices \cite{CastroRMP2009,DasSarmaRMP2012}.
Due to the large optical phonon energy of $\Omega\approx200\;\mathrm{meV}$ in graphene, the emission of acoustic
phonons is the only efficient cooling pathway for hot electrons below the energy $\Omega$.
Unlike conventional metals with large Fermi surfaces, where the Debye temperature sets the boundary between high- and low-temperature behavior in the electron-phonon interaction,
a new characteristic temperature arises in graphene, the Bloch-Gr\"{u}neisen temperature ($T_{\text{BG}}$).
It results from the small Fermi surface in graphene and is defined by the maximum phonon
wave vector $q_{max} = 2 k_F$ (Fig.\,\ref{fig:illustration}b).
Above $T_{\text{BG}}$, only a fraction of acoustic phonons with wavevector $q \leq 2 k_F$ can scatter off electrons.
While this phase space restriction leads to the crossover behavior of electron-phonon resistivity between high-temperature $\rho(T) \propto T$ and low-temperature $\rho(T) \propto T^{4}$ dependence \cite{Efetov2010PRL},
it also imposes a significant constraint on the cooling of hot electrons; the energy dissipated by acoustic phonons cannot exceed $k_B T_{\text{BG}}$ per one scattering event (Fig.\,\ref{fig:illustration}c).
As a result, the electron-lattice cooling can be slow for phonon temperatures $T_{ph} \gg T_{\text{BG}}$, requiring many scattering events to dissipate the heat of the hot electrons.

On the other hand, a recent theory predicts alternative cooling pathways mediated by \emph{supercollisions} \cite{Song2012PRL}.
Here, disorder-assisted phonon scattering or two-phonon scattering events enable the emission of acoustic phonons with wave vectors $q \geq 2 k_F$,
thus transferring higher energy than normal collisions. Such collisions occur less frequently than ordinary collisions,
but dominate the electron-lattice cooling by utilizing the entire thermal distribution of phonons (Fig.\,\ref{fig:illustration}c).

\begin{figure}[b]
\includegraphics[width=7.8cm]{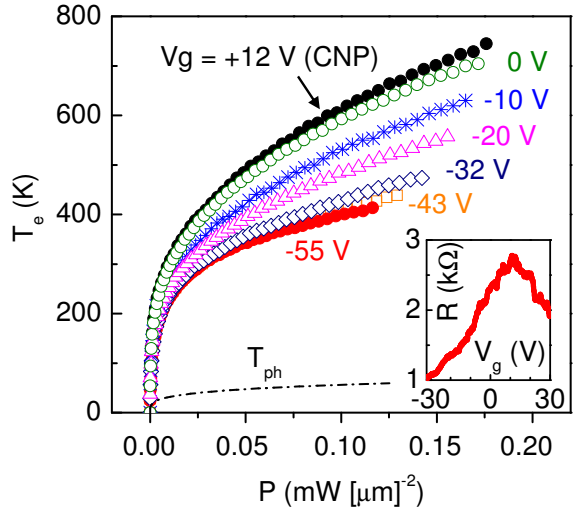}
\caption{ \textbf{Electron temperature in
graphene as a function of supplied Joule power.}
Electron temperature as a function of Joule power per unit area, $P=V^2/(R\,WL)$, in a sample of width $W= 2.8$ $\mu\text{m}$ and length $L = 2.2$ $\mu\text{m}$.
Data is shown for selected gate voltages: $V_{\text{g}}$ ranges from high carrier density (-55~V) to charge neutrality ($V_{\text{g}} = 12$~V).
For comparison, the lattice temperature (dashed line) is estimated from $T_{ph}\approx(P/\Sigma_K)^{1/4}$ with the lattice-substrate coupling constant $\Sigma_K\simeq 10 \;\mathrm{Wm^{-2}K^{-4}}$.
Inset: Resistance as a function of gate voltage at 4.2~K.
}\label{fig:Te{P}}
\end{figure}

A unique feature of graphene is the tunability of $T_{\text{BG}}$ with Fermi energy $E_{\text{F}}$ \cite{Efetov2010PRL}.
As both the electron and the phonon energy are linear in wave vector, the Bloch-Gr\"{u}neisen temperature is simply given by
$k_B T_{\text{BG}} = (2 v_s / v_F) E_{\text{F}}$, where $v_s$ and $v_F$ are the sound and the Fermi velocity, respectively.
Hence, both energy scales, $k_B T_{\text{BG}}$ and $E_{\text{F}}$, are linked by the ratio $2 v_s / v_F \approx 0.04$.

In this work, we experimentally test the different models
of electron-phonon cooling \cite{Kubakaddi2009PRB,Bistritzer2009PRL,Tse2009PRB,Viljas2010PRB,Song2012PRL}.
The low- ($T_{ph} < T_{\text{BG}}$) and high-temperature ($T_{ph} > T_{\text{BG}}$) regimes become accessible by
tuning the size of the Fermi surface, i.e.~tuning $T_{\text{BG}}$ by means of an electrostatic gate potential $V_{\text{g}}$ (Fig.\,\ref{fig:illustration}).
Electrons are heated with a bias voltage $V$ in a two-probe configuration, and the electron-lattice cooling rate is investigated
by means of Johnson noise thermometry, a primary temperature measurement technique based on the fundamental properties of thermal fluctuations in conductors.
The electron temperature is deduced from radio frequency shot noise measurements,
relying on the relation $S_I=4k_BT_e/R$ between the current noise spectrum $S_I$, the sample resistance $R$, and  $T_e$.
Johnson-noise thermometry has proven useful to study carbon nanotubes \cite{Chaste2010APL,Wu2010APL} and more recently graphene \cite{Betz2012PRL,Fay2011PRB}.
In the high-temperature regime $T_{ph} \geq T_{\text{BG}}$, we find an energy relaxation rate,
$J_{\text{SC}}= A \,(T_e^{3}\, - \, T_{ph}^{3})$, as predicted by the supercollision mechanism \cite{Song2012PRL},
with a prefactor $A$ related to the amount of disorder and the carrier density.
Uniquely in graphene, there is a regime where disorder-assisted SCs dominate over the conventional electron-phonon collisions \cite{Kubakaddi2009PRB,Bistritzer2009PRL,Tse2009PRB,Viljas2010PRB} above $T_{\text{BG}}$.
In the low-temperature regime, for $T_{ph} < T_{\text{BG}}$, we regain the dependence $J \propto (T_e^{4}\, - \, T_{ph}^{4})$ \cite{Kubakaddi2009PRB,Viljas2010PRB,Betz2012PRL,Baker2012PRB},
which is the signature of standard electron-phonon interaction in 2D graphene.

Figure~\ref{fig:Te{P}} shows the increase of electron temperature as a function of dissipated Joule power in a monolayer graphene device on a hexagonal boron nitride (h-BN) substrate.
The average electron temperature $T_e$ is extracted from the shot noise $S_I$, measured in the MHz to GHz band to overcome the environmental $1/f$ contribution (see Methods).
With the application of the Joule power $P$ to the electron system, $T_e$ rises well above the substrate temperature $T_{\text{0}} = 4.2$~K:
$T_e$ reaches 300--700 K when $\sim$1 $\mathrm{mW}$ of Joule power is generated in our $\sim$5 $\mu\mathrm{m}^2$ sized sample.
The large increase of $T_e$ is, partly, due to the much smaller carrier density in graphene compared to a conventional metal.
The Joule power is dissipated in a relatively small number of carriers, resulting in the corresponding increase of  $T_e$.
Accordingly, we find $T_e$ to be at its largest value at the charge neutrality point (CNP), located at $V_{\text{g}} = 12$~V (inset of Fig.~\ref{fig:Te{P}}).
Moving away from the CNP (i.e.~increasing the carrier density $n_{\text{s}}$) $T_e$ decreases at constant power $P$.

The large thermal decoupling of electrons and phonons ($T_e\gg T_{ph}\gtrsim T_{\text{0}}$) reflects the weak electron-phonon interaction in graphene.
The steady-state value of $T_e$ is determined by the balance between the Joule heating $P$ and the cooling powers $J$ at play\,\cite{Wellstood1994PRB}.
The cooling of electrons occurs via either  heat transfer to the phonons or heat diffusion to the metallic leads.
At sufficiently high bias the diffusive contribution can be neglected,
and the electron-lattice cooling can be directly investigated \cite{Betz2012PRL}.
Similarly, $T_{ph}$ reaches a steady-state value when the same power from the electrons is transferred to the substrate with a typical black-body radiation law $P=\Sigma_K(T_{ph}^4- T_{\text{0}}^4)$.
The phonon-substrate coupling constant \cite{Balandin}, $\Sigma_K\simeq 10 \;\mathrm{Wm^{-2}K^{-4}}$, is $3$--$4$ orders of magnitude larger than the electron-phonon coupling constant in graphene \cite{Betz2012PRL},
and we achieve phonon temperatures $T_{ph}\simeq 4$--$65\;\mathrm{K}$ with $P=0$--$0.2\;\mathrm{mW/\mu m^2}$.

\begin{figure}
\includegraphics[width=8cm]{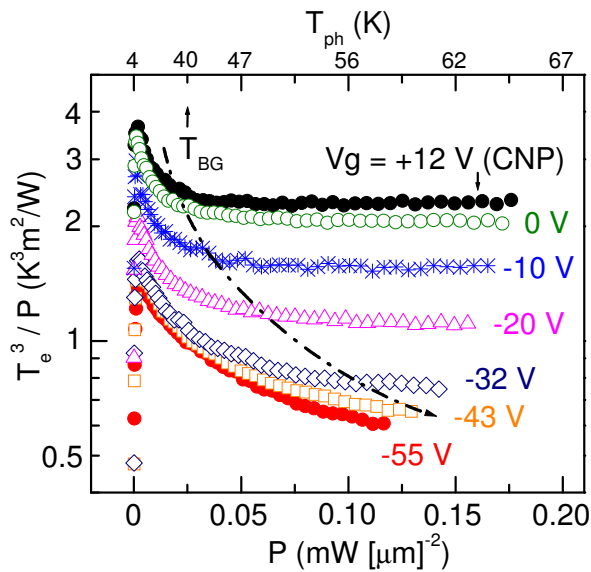}
\caption{ \textbf{Supercollision cubic law.} Electron temperatures are plotted as $T_e^3/P$ for a range of selected gate voltages.
$T_{BG}$ is tuned by the electrostatic gate potential, and displayed by the dashed-dotted line.
For temperatures $T_{ph} > T_{\text{BG}}$, the SC cooling dominates, visible from the arising plateaus in the $T_e^3/P$ representation.
}\label{fig:T3}
\end{figure}

In order to test the supercollision mechanism, we plot in Fig.~\ref{fig:T3} the electron temperature in the representation $T_e^3/P$ as function of $P$. SC theory predicts the energy loss power
\begin{equation}
J_{\text{SC}}= A \,(T_e^{3}\, - \, T_{ph}^{3}),\;\;\;\;\;\;\; A = 9.62 \,\frac {g^2\nu^2(\mu)k_B^3}{\hbar k_F l},
\label{SCequation}\end{equation}
while modeling disorder by short-range scatterers.
Here $g$ is the electron-phonon coupling, $\nu$ is the density of states per spin/valley flavor, and $l$ is the mean free path.
In this model $k_F l$ is a constant inversely proportional to the strength and concentration of impurities.
Other types of disorder like ripples \cite{Song2012PRL} or to some extend long range impurities can give rise to a similar $T^3$ dependence with however different expressions for the coupling constant.
Note that, with $T_e \gg T_{ph}$, the $T_e^3/P$ plateau corresponds to the inverse of the coupling constant $A$.

The $T_e^3/P$ plots (Fig.~\ref{fig:T3}) demonstrate three different cooling regimes;
the dip at $P\approx0$ representing the heat conduction to the leads,
the asymptotic $P\propto T_e^3$ behavior (the $T_e^3/P$ plateau) pronounced in the high-temperature regime ($T_{ph} > T_{\text{BG}}$),
and the low-temperature $P\propto T_e^4$ behavior \cite{Betz2012PRL} at $T_{ph} < T_{\text{BG}}$ which translates into a decreasing function $T_e^3/P\sim P^{-1/4}$.
In order to highlight the border between the low and high temperature regimes,
the gate-tuned $T_{\text{BG}}$ is visualized by a dashed-dotted line which connects the values of $T_{\text{BG}}$ at each $V_g$,
while $T_{ph}$ is denoted in the upper x-axis of the Fig.~\ref{fig:T3}.
We use the relation
$T_{\text{BG}} = (2 v_s / v_F)E_{\text{F}}/k_B \approx 54\sqrt{n_s}\;\mathrm{K}$
where the carrier density  $n_s=\sqrt{[C_g(V_g-V_{\text{CNP}})/e]^2 + n_0^2}$  (in  units of $10^{12}$~cm$^{-2}$) is deduced from gate voltage $V_g$ using
the gate capacitance $C_g = 35 \;\mathrm{aF/\mu m^2}$. $C_g$ is estimated from device geometry.
We have introduced a residual density $n_0 \simeq 4\times 10^{11}$~cm$^{-2}$ to account for the minimum conductivity due to electron-hole puddles at the CNP.

Our observation of the $T_e^3/P$ plateaus, which are most pronounced near the charge neutrality, strongly supports the SC cooling mechanism in the high-temperature regime.
With increasing $n_s$ (resp.  $T_{\text{BG}}\propto\sqrt{n_s}$),  plateaus start to develop at larger $P$ (resp. $T_{ph}$), up to a point where
the low-temperature regime ($T_{ph} < T_{\text{BG}}$) dominates over the entire power range (for $V_g \lesssim -32$~V).

\begin{figure}[b]
\includegraphics[width=8cm]{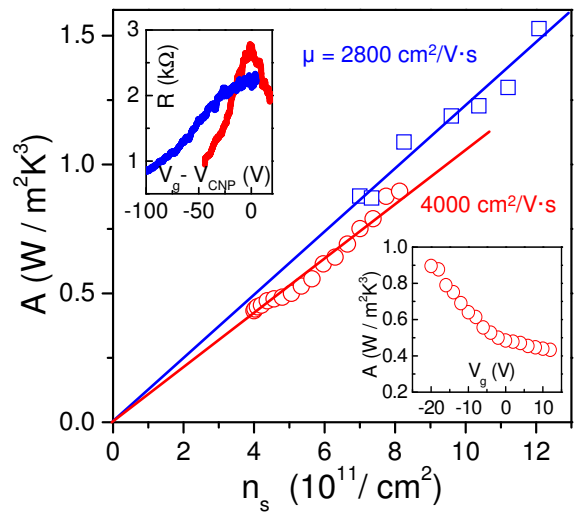}
\caption{ \textbf{SC Coupling constant $A$ as a function of carrier density.}
The coupling constant $A$ increases linearly with the carrier density and is dependent
on the level of disorder in the sample.
The latter is changed by a heat treatment: red dots correspond to data acquired before and blue squares to measurements after the annealing.
Right inset: Coupling constant $A$ as a function gate voltage, deduced from the $T_e^3/P$ plateaus for steps of 2~V in the
range of -20~V $\leq$ $V_g$ $\leq$ 12~V.
Left inset: Resistance versus ($V_g - V_{\text{CNP}}$) measured at 4.2~K before (red) and after (blue) the heat treatment.
}\label{fig:coupling_constant}
\end{figure}

We now focus on the high-temperature regime and compare in greater detail the SC theory to our data.
The short-range disorder-assisted SCs, shown in Eq.~(\ref{SCequation}), scale linearly with carrier density, $n_s \sim \nu^2(\mu)$.
In Fig.~\ref{fig:coupling_constant}, the SC coupling constant $A$, extracted from the $T_e^3/P$ plateau, is presented for $-20 \;\text{V} \leq V_g \leq 12 \;\text{V}$ (red dots).
Indeed, we find a linear dependence $A \propto n_s$, in a good agreement with the SC theory.
Evaluating Eq.~(\ref{SCequation}), we obtain $A \approx 7.5\times 10^{-4} \; \frac {D^2 n_s}{k_F l}\;\,[\text{W/(m$^2$K$^3$)}]$,
where $D$ is the deformation potential in the unit of eV, and $n_s$ in the unit of $10^{12}$~cm$^{-2}$.
Taking $k_Fl = \sigma \frac {h}{2e^2} \approx 3.5$, from conductivity data, the slope in Fig.~\ref{fig:coupling_constant}
  indicates a deformation potential $D \approx 70 \; \text{eV}$, larger than the reported values $D\sim 10-30\; \text{eV}$ \cite{Efetov2010PRL,Bolotin2008PRL,Dean2010natnano}.
  We are led to the conclusion that
 although Eq.~(\ref{SCequation}) gives the order of magnitude, correct carrier density, and temperature dependence of the cooling power, it also leads to an underestimate of the coupling constant.
 This is possibly an indication that the short-range scattering hypothesis is too restrictive.

In order to check the disorder dependence in $A$,  we have increased the impurity content of the sample by a heat treatment;
this has shifted the CNP to 115~V, and decreased the mobility accordingly (see left inset of Fig.~\ref{fig:coupling_constant}).
Consistently we observe a $\sim30 \%$ increase of the the coupling constant (blue squares in Fig.~\ref{fig:coupling_constant}), which confirms
the effect of impurity concentration as predicted in Eq.~(\ref{SCequation}).

Turning to the discussion of the results,
the enhancement factor for the disorder-assisted SC cooling over the conventional cooling pathways, $J_{\text{0}}\propto(T_e -  T_{ph})$, is given as in Ref.~\onlinecite{Song2012PRL}
\begin{equation}
\frac {J_{\text{SC}}} {J_{\text{0}}} = \frac {0.77} {k_Fl} \, \frac {T_e^{2}+T_eT_{ph}+T_{ph}^{2}} {T_{\text{BG}}^{2}}.
\label{SCequation2}\end{equation}
At the CNP, we have $T_e \approx 400 \,\text{K}$ and $T_{ph} \sim T_{\text{BG}} \approx 40\,\text{K}$ for $P= 0.025 \;\mathrm{mW/\mu m^2}$  (Figs.~\ref{fig:Te{P}} and~\ref{fig:T3}).
The enhancement factor $J_{\text{SC}}/J_{\text{0}}$ is as large as 80 times, and explains the immediate observation of the SC cubic law right above the $T_{\text{BG}}$.
Note that with further increase of the $P$ ($T_e$ and $T_{ph}$), the enhancement factor becomes even larger.

While the SCs cubic behavior and the linear law for normal collisions are valid in the degenerate limit, $k_BT_e < E_F$,
the normal electron-phonon collisions are predicted to give $J \propto T_e^{4}\,(T_e - T_{ph})$ for $k_BT_e \gg E_F$ \cite{Viljas2010PRB,Bistritzer2009PRL}.
In our experiment, the residual carrier density at the CNP limits $E_F$ above 65~meV, and we could not achieve the regime $k_BT_e \gg E_F$.
Higher mobility samples are needed to access this non-degenerate regime as well as electron cooling by optical phonons which are both of great interest.

In conclusion, we have investigated impurity mediated electron-phonon interaction in diffusive graphene by measuring the energy-loss of hot-electrons both below and above the Bloch-Gr\"{u}neisen temperature for phonons. We observe the $T_e^3$ dependence on the electronic temperature predicted for supercollisions at high temperature while recovering the $T_e^4$ dependence for ordinary collisions at low temperature. The Bloch-Gr\"{u}neisen crossover temperature agrees with the estimated phonon temperature and the coupling constant is consistent with the carrier density and disorder dependence predicted for short-range impurity scattering.
Beside its implication for electron-phonon physics, our work is of direct relevance for the performance of graphene bolometers and photo-detectors \cite{Gabor2011Science,Vora2011arXiv,Yan2011arXiv,Fong2012arXiv}.

During the writing of this manuscript, we have become aware of the preprint \cite{Graham2012arXiv}  dealing with the SC cooling in graphene.
In this experiment the photocurrent generated at a graphene p-n junction is well described by the SC cubic law.

\section{Methods}
The experiments have been performed on exfoliated monolayer graphene on h-BN/SiO$_{2}$/Si substrates.
The h-BN platelet was exfoliated from a high quality h-BN powder (St. Gobain "Tr\`es BN".)
The graphene flake was subsequently placed on top of the h-BN using a transfer technique described
in Ref.~\onlinecite{TransferDelft}.
The heavily p-doped Si was used as a back gate and the thickness of the insulating layer was $\sim$$1\;\mathrm{\mu m}$.
The samples, produced by means of e-beam lithography and dry etching, were embedded in a coplanar wave-guide adapted for GHz frequencies\cite{Pallecchi2011APL}.

We applied a Joule power $P=V^2/R$ to create the hot carrier population and study the electronic cooling as shown in Figs.~\ref{fig:Te{P}} and \ref{fig:T3}.
The sample's length and width were $L=2.2\;\mathrm{\mu m}$ and $W=2.8\;\mathrm{\mu m}$, respectively.
The sample resistance $R\sim1/n_s$ was calculated from the voltage drop across the bias resistance $R_{bias}= 4.7 \;\mathrm{k} \Omega$ (see Fig.~\ref{fig:illustration}d).
The carrier density $n_s$ was controlled electrostatically by means of the back-gate voltage $V_g$.

All data was measured for samples in a direct contact with liquid helium to ensure a cold phonon bath.
The noise signal resulting from the hot carriers was first amplified by a cryogenic low noise amplifier at $T = 4.2\;\mathrm{K}$ followed by two additional amplifiers at room temperature (overall gain of $\approx 82 \;\mathrm{dB}$).
A $460\;\mathrm{Ohms}$ Al/AlOx/Al tunnel junction was used to calibrate the shot noise.
Our broad-band setup, with a bandwidth of $\sim$$1\;\mathrm{GHz}$, allows us to quantitatively separate the white shot noise contribution $S_I$ from $1/f$ noise by fitting the spectra with a $S_I+ C/f$ law.
We find $C\propto I^2$ in accordance with the Hoodge law \cite{Hoodge1981RPP}. This procedure is especially important for small samples and high currents where the $1/f$ contribution can be large.

\begin{acknowledgments}
The research has been supported by the contracts ANR-2010-BLAN-MIGRAQUEL, SBPC and Cnano-2011 Topin's and Grav's.
\end{acknowledgments}

\end{document}